\documentclass[12pt]{revtex4}
\usepackage{graphicx}
\usepackage{color}

\newcommand{\sub} [2] {\mbox{$#1_{\mbox{\scriptsize{#2}}}$}}

 \newcommand {\beq}{\begin{equation}}
\newcommand {\eeq}{\end{equation}}

\begin{document}

\begin{titlepage}
\centerline{\tt Zimmer-PC:F:/Neil/project documents/Papers/14\_1 Redef redux/text/14\_10 redef educational proposal}
\vspace{.2in}
\centerline{\large\bf "What is The SI?" A Proposal for an Educational Adjunct to the SI Redefinition}
\vspace{.1in}
\centerline{Neil M. Zimmerman and David B. Newell}
\vspace{.2in}
\centerline{+1-301-975-5887}
\centerline{neilz@mailaps.org; ftp://ftp.nist.gov/pub/physics/neilz/papers.html}
\centerline{National Institute of Standards and
Technology\footnote{Quantum Metrology Division, Physical Measurements Laboratory,
U.S. Department of Commerce.  Official
contribution of the National Institute of Standards and Technology,
not subject to copyright in the United States.}}
\centerline{Gaithersburg, MD 20899, USA}
\vspace{.2in}
\centerline{\today}

\begin{abstract} 

We discuss how the likely 2018 redefinition of the SI system of units might affect the
ability of students to understand the link between the units and the new system.  The
likely redefinition will no longer define a set of base units, but rather a set of
constants of nature, such as the speed of light $c$ and a particular hyperfine splitting
in Cs $\Delta\nu(^{133}$Cs)$\sub{}{hfs}$.  We point out that this list of constants need
not be the only way to introduce students to the subject, either in class or in textbooks.
We suggest an alternative way to introduce high school and undergraduate students to the
redefined SI, by suggesting a list of experiments for some units; this list would be
completely compatible with the redefined SI, and would have all of the same scientific and
technological advantages.  We demonstrate by questionnaire results that this alternative
is more appealing to students.  We hope to spur a discussion amongst teachers regarding
this important topic for high school and undergraduate physics courses.

\end{abstract}

\maketitle
\end{titlepage}

%
%

%

The system of units (SI or Syst\`{e}me International d'Unit\'{e}s) is likely to be
fundamentally changed in 2018\cite{CIPM}\cite{Newell14a}\cite{Aubrecht12a}.  One aspect of
the fundamental change is that the SI will no longer be defined in terms of experiments
that lead to the units, but rather with respect to a set of fundamental constants of
nature.  We are concerned about the effect of this change on the education of beginning
(senior high school and early college) science, technology, engineering, and mathematics
(STEM) students, because we believe it is easier for students early in their career to
grasp the concept of units as being based on experiments.

Before we jump into a detailed discussion of the redefinition of the SI, we will give a
general introduction into the need for any system of units, and how the present SI
evolved.  It could be argued that the need for physical measurements have existed at least
since we had to judge where to place one foot in front of the other in a controlled fall
known as walking.  Indeed, we perform measurements all the time subconsciously as we
estimate or judge physical quantities surrounding us such as length, size, weight, or
speed.  With the development and advancement of the first ancient agricultural societies
came the need to standardize physical measurements to facilitate trade and commerce.  It
is interesting to note that at the foundation of these early measurement systems was a
``universal measure'' that was available to everyone, anytime, anywhere, such as a seed or a
grain, from which measurement units related to length, area, volume and mass were
constructed \cite{Williams14a}.

The existing SI is also based on perceived universal measures dating back to the formation
of the metric system during the French revolution.  Back then the Earth's meridian was
chosen as a constant of nature or a universal measure from which an everyday practical
unit of length, the meter, was based\cite{Adler02a}.  Derived from the meter was the
practical unit of mass, the kilogram, based upon a certain volume (0.001 m$^3$) of water
(distilled at 4 $^\circ$C).  Together the meter, kilogram, and second (being based on
astronomical observations) provided a practical system of units not only for measuring
everyday physical quantities used in trade and commerce, but also for measuring newly
discovered phenomena such as electricity and magnetism.

Today, the present SI defines explicitly seven base units, of which the most familiar are the meter,
kilogram, and second.  Not only are they are used for everyday practical measurements,
but, together with simple dynamics and Newtonian physics, they are used as educational
tools for introducing the system of units.  In addition, with the exception of the
kilogram\cite{Pratt14a}, the definitions of the other six present-day SI base
units have the effect of implicitly defining six true invariants of nature, such as the
hyperfine splitting frequency of caesium atoms $\Delta\nu(^{133}$Cs)\sub{}{hfs} and the
speed of light $c$.

There is an international effort to redefine the SI so that it defines explicitly seven
true invariants of nature\cite{CIPM}\cite{Newell14a}\cite{Aubrecht12a}.  This will have the
effect of implicitly defining a large number of possible experiments which could be used
to realize the various units.  The basic
motivations for the change can roughly be distilled as: i) replacing the only remaining
artifact standard (i.e., a standard based on a particular piece of metal located in Paris,
France), the International Prototype of the Kilogram (the IPK), with a definition of mass
based on fundamental laws of physics; ii) using the excellent reliability and
reproducibility of the quantum electrical standards; iii) defining the SI explicitly in
terms of fundamental constants (the ``defining constants'') rather than base units.

We note several excellent resources for detailed discussions of this variety of motivations:
For discussions of mass including the present system and the possible experiments in the
new SI (the ``electronic kilogram'' and a nearly-perfect Si sphere), please see
Ref. \cite{Pratt14a}.  For a discussion of the system of electrical units including the
``quantum electrical standards'' (Josephson voltage and quantum Hall resistance
standards), please see Ref. \cite{Fletcher14a}.  Finally, a discussion of the physics of
the fundamental constants, and how the best values are periodically achieved, is in
Ref. \cite{Mohr10a}. 

While the motivations and the practicality of the redefinition seem to us to be quite
compelling for practitioners in metrology (the science of measurement), we believe that
presenting the new definition of the SI strictly in terms of exact values of fundamental
constants [motivation iii)] may not be the best way to introduce the topic of the new SI
to students.  There are several ways that we can make this point: 1) The "SI" has the word
units in it, and many students may find that a definition of the System of Units which has
no units named in it confusing\cite{Aubrecht12a}; 2) As pointed out by one of
us\cite{Newell14a}, it may be quite challenging for teachers of senior high school students
or beginning college students to explain how the unit of mass flows from the value of
Planck's constant $h$, which is typically not introduced until much later in a student's
career; 3) Similarly, many physics textbooks have a discussion of or an appendix on units,
and explaining concisely how (for example) mass depends on $h$ (see definition C in
Fig. 2) may again be quite challenging.

There are many advantages to the likely redefinition of the SI, including i) replacing the
IPK, ii) having a system of units which depends only on fundamental laws and invariants of
nature; iii) making it possible for all units (not just the seven base units) to be
expressed in terms of seven defining constants which are defined to have zero uncertainty
[for example, in temperature the SI unit Kelvin would be 1 K = 2.266 665 ($h
  \Delta\nu(^{133}$Cs)$\sub{}{hfs} / k$), where $k$ is the Boltzmann constant] ; iv) eliminating 
the "1990 conventional electrical units", which refers to the fact that for the past 25
years, electrical metrologists have used the quantum electrical standards (Josephson
voltage standard and quantum Hall resistance standard)\cite{Fletcher14a} to disseminate
electrical standards, which means that they are not reporting their results in the SI
units\cite{Zimmerman98a} but rather in the so-called ``1990 units''; v) by virtue of the
seven defined constants, allowing scaling of definitions of units over many orders of
magnitude.

In the proposal contained in this paper, we suggest that students be first introduced to
the SI in terms of units rather than defined constants.  We wish to emphasize that this
proposed educational adjunct to the SI is completely compatible with the likely 2018
redefinition; in particular, all of the advantages listed above for the likely
redefinition are also reflected in our proposed introduction based on units.  We also wish
to emphasize that our proposal is not for a definition of the SI, but rather for the use
of teachers in introducing the redefined SI to students, either in classes or in
textbooks.

\section{Three Possible Introductions to the SI}

\begin{figure}[htbp] 
\includegraphics[trim = 30mm 140mm 20mm 25mm, clip, scale = 0.7]{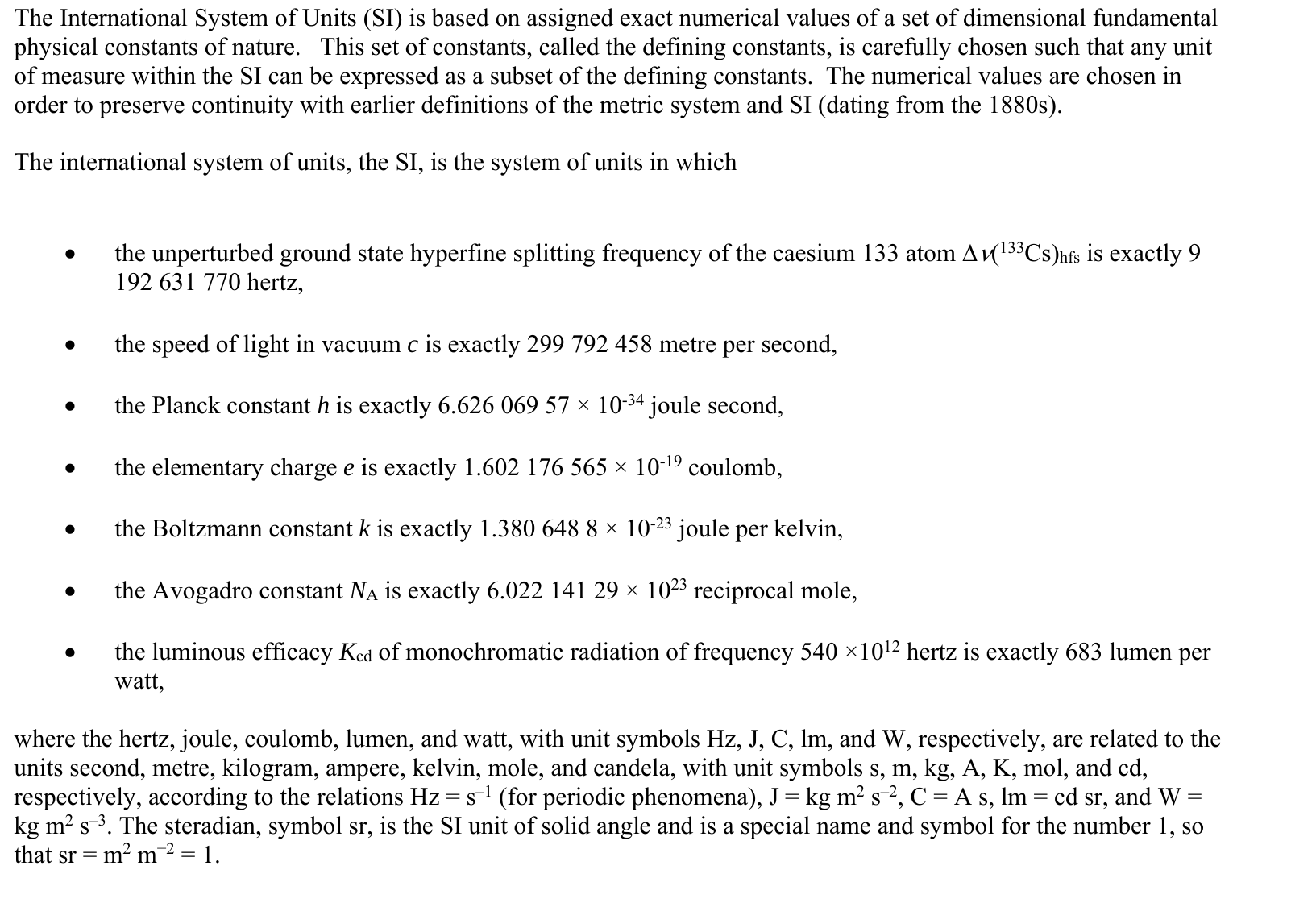}
\caption{\label{fig:prop_A} Proposal A: Defining Constants.  From Section 2.2 of Dec 2013 draft of the 9th SI Brochure\cite{CIPM}}
\end{figure}

\begin{figure}[htbp] 
\includegraphics[trim = 30mm 50mm 20mm 25mm, clip, scale = 0.7]{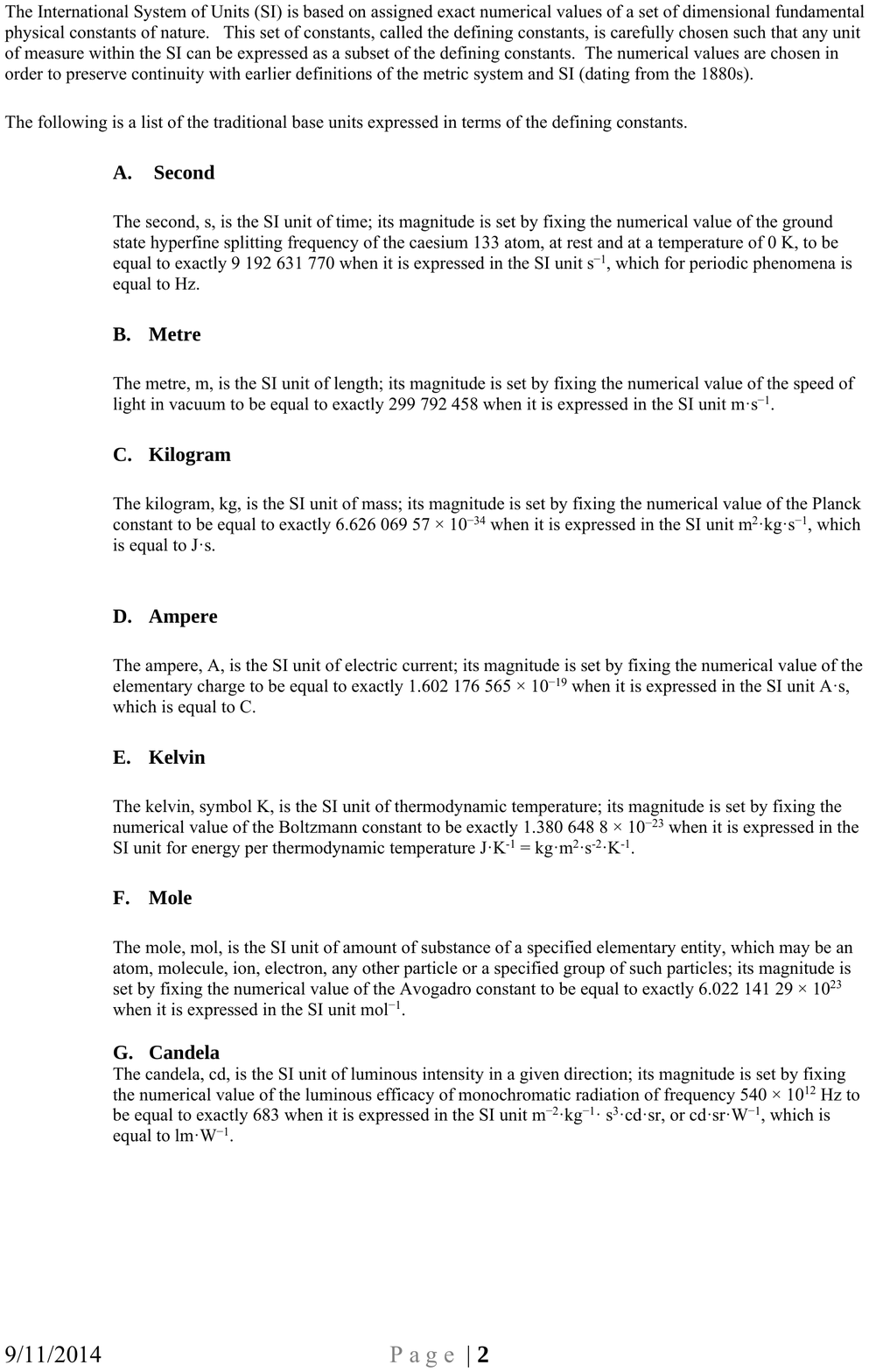}
\caption{\label{fig:prop_B} Proposal B: Traditional Base units expressed in terms of defining Constants. Adapted from Section 2.4 of Dec 2013 draft of the 9th SI Brochure\cite{CIPM}}
\end{figure}

\begin{figure}[htbp] 
\includegraphics[trim = 30mm 60mm 20mm 25mm, clip, scale = 0.7]{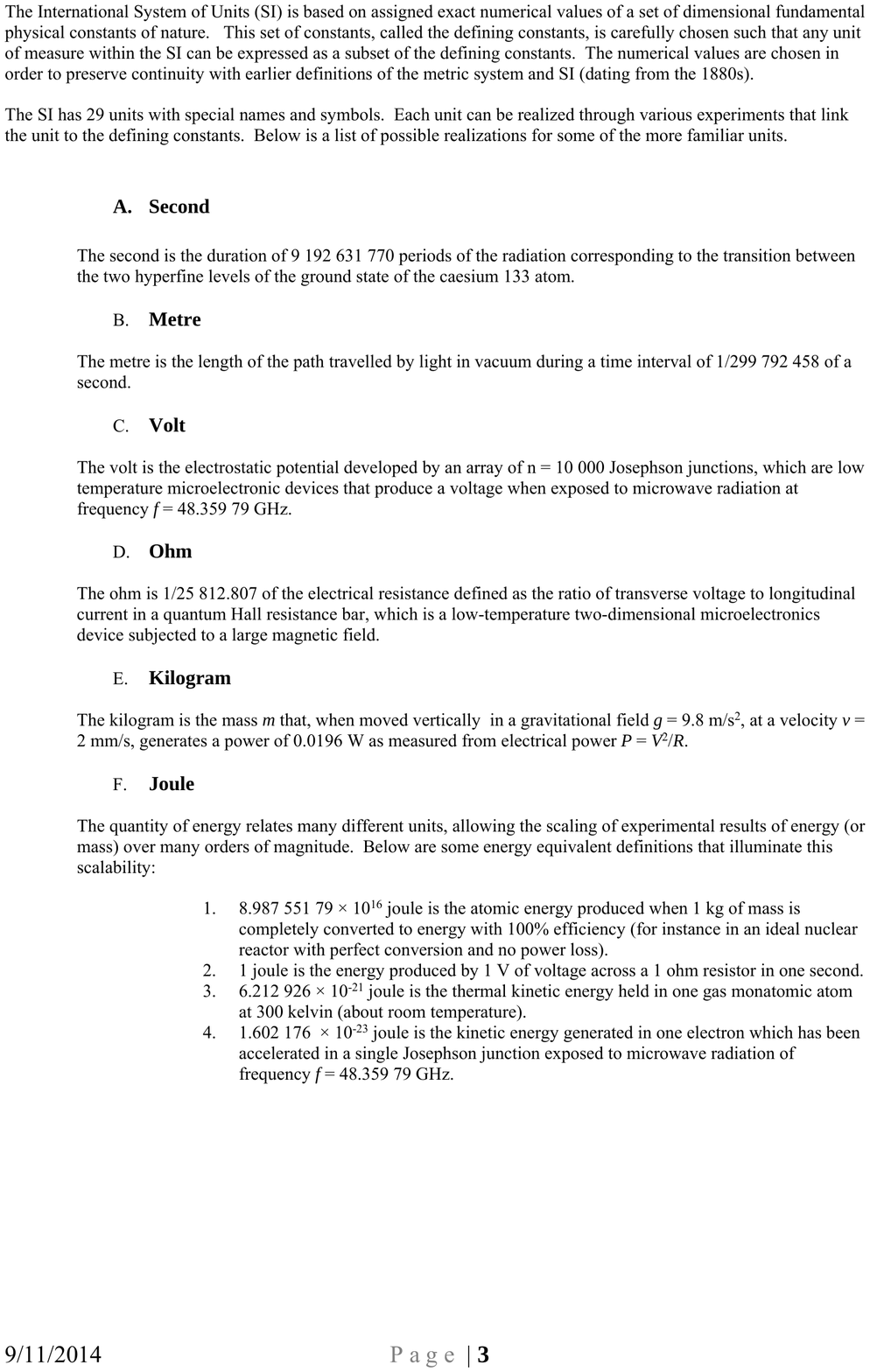}
\caption{\label{fig:prop_C} Proposal C: Definitions Based on Experiments}
\end{figure}

We refer to Figures \ref{fig:prop_A}, \ref{fig:prop_B}, \ref{fig:prop_C}; these contain
three possible versions of the text which could be used to introduce students to the
subject of units.  Briefly, Proposal A is quoted from the draft ninth edition of the SI
brochure\cite{CIPM} [from here on we refer to this as the ``new SI brochure''], and sets
out the definitions and values of the seven defined constants; Proposal B is adapted from
a different Section of the new SI brochure, and lists the seven traditional base units as
defined in terms of the seven constants; Proposal C is an alternative proposal, and
contains a list of what we believe to be the more commonly-used units, defined not in
terms of constants but in terms of brief descriptions of experiments which could be used
to realize those units.  As discussed at the end of the Introduction and Motivation, we
note that these three Proposals are interchangeable in terms of both the effect on the
system of units, and on satisfying all the advantages listed above.  In all three
Proposals, we added the same introductory paragraph for the purposes of the questionnaire;
this paragraph does not appear in the new SI brochure.

We now discuss how we derived these three Proposals: Proposal A is quoted from the new SI
brochure\cite{CIPM}, and we suspect is what most metrologists who have been involved in the
likely redefinition would answer when asked: "What is the SI?".  Proposal C is an attempt
to address the motivation that we have discussed, in terms of introducing students to the
units based on simple descriptions of the types of experiments that one might use to
realize those units.  We discuss how we arrived at this Proposal in more detail directly
below.  

As with Proposal A, Proposal B is taken from a different section of the new SI brochure
and was thus devised by the metrologists who are advocating the likely redefinition.  It
has a list of units, but the definitions are based on invariants of nature and not on
experimental realizations; it thus reflects some of the attributes of the two other
Proposals.

Turning in more detail to Proposal C, we will note that we took this Proposal through
several versions.  In an earlier, longer version of this Proposal, we had much more
detail.  Although the longer Proposal C was favored by students in a questionnaire
comparing their understanding of the motivation underlying each Proposal (see Appendix), a significant
number of them also commented that the original Proposal C was too long and complicated.
We thus pared out a lot of the detail including all of the mathematical formulae
converting the unit to the appropriate combination of defined constants.  The resulting
Proposal C is the one included in Figure \ref{fig:prop_C}, and the one that we propose as
a starting point for discussion in the community.

In addition, we also note that we have included a paragraph discussing different ways of
realizing measurements of energy.  We included this for two reasons that we believe appeal
to students: 1) It manifestly demonstrates that the relationships between various
quantities in both the present and the proposed SI are largely governed by the requirement
that energy derived in different ways (e.g., thermal, mechanical, mass-equivalent,
quantized) must be coherent; 2) It also manifestly demonstrates the remarkable ability of
the SI to allow measurements of the same quantity that vary by many many orders of
magnitude.

\section{Questionnaire Details and Results}

We asked students to respond to an informal questionnaire at the University of Maryland, College Park.  This questionnaire was given to students in a recitation section of the introductory physics sequence for physics students; students were free to attend or not attend the section, and about half of the student body (29/55) chose to attend.  These students read the three proposals and give us their numerical ratings and qualitative comments; please see Figure \ref{fig:cover_page} for the wording of the questionnaire.

The students were roughly equally split by gender, with 80\% studying Physics, and the rest from a variety of technical and non-technical disciplines.  About half were freshmen, one third had finished one year of undergraduate studies, 15 \% had finished two years, and the rest more than two years.

We can quote two written comments that we found quite illuminating, in terms of relative ease of understanding between the three Proposals:

\begin{itemize}
\item ``I preferred proposal C.  This proposal was clear and expressed the SI units in terms of conceivable phenomena/measurements whereas A associated the SI units with less conceivable (however reasonable) phenomena/measurements.  While proposal A was clear, it was not as helpful in connecting the quantitative aspects of the units with the qualitative aspects, i.e., the phenomena the units are measurements of.''

\item ``I felt that the two best proposals in terms of contents were B and C.  Even though they were describing similar things using constants such as the speed of light to describe a meter, I felt proposal C did the best job.  It actually put a new value to the common SI units, which I feel is more effective than just saying what constant the SI unit relates to (seen in proposal B).  If the point of this is to completely redefine SI units, proposal C does a much better job conveying the information and the basic theme of the set of definitions was much clearer.  As for proposal A, I felt that even though the wording was a lot easier to understand, it didn't fully convey the new information and was just too simplistic.''

\end{itemize}

\begin{figure}[htbp] 
\includegraphics[trim = 0mm 0mm 0mm 0mm, clip, scale = 0.5]{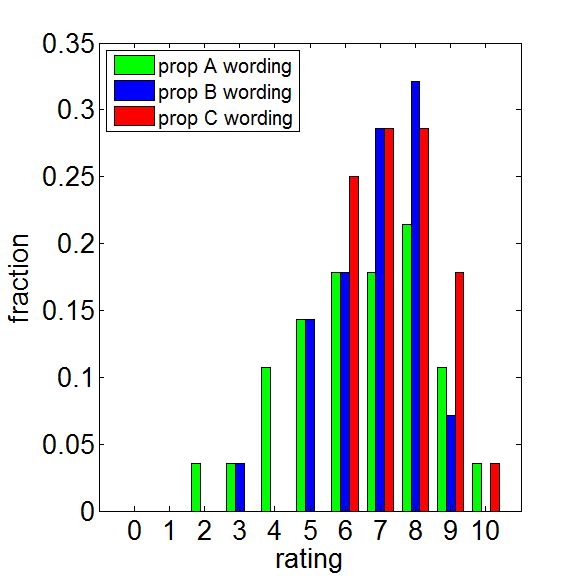}
\caption{\label{fig:questionnaire_wording} Questionnaire Results: Wording}
\end{figure}

\begin{figure}[htbp] 
\includegraphics[trim = 0mm 0mm 0mm 0mm, clip, scale = 0.5]{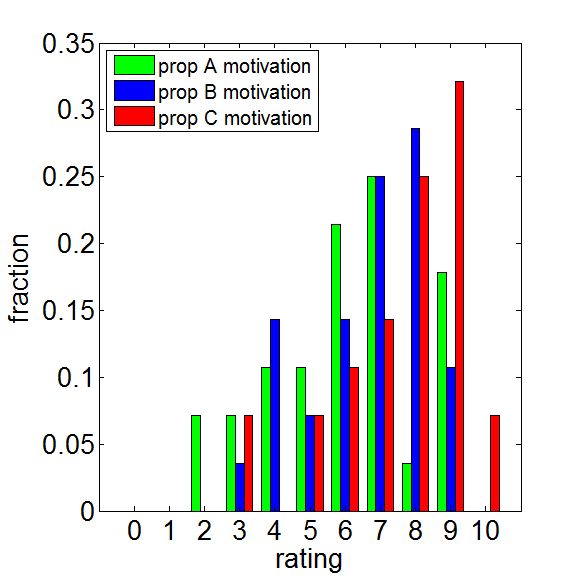}
\caption{\label{fig:questionnaire_motiv} Questionnaire Results: Motivation}
\end{figure}

We see from both Figures \ref{fig:questionnaire_wording} and
\ref{fig:questionnaire_motiv}, and from the Table, that there were significant
differences in the students' rating of both i) the clarity of the wording in the three
Proposals and ii) how compelling the motivation or conceptual underpinning was.  Both the
Figures and the numerical values in the Table make it clear that the students modestly preferred
Proposal C over B, and both substantially over A.

As discussed earlier, we started with a much longer and more detailed Proposal C; we also presented the comparison of A, B, and original C in an earlier questionnaire to 55 students who came from a much broader range of disciplines.  The ratings in this earlier questionnaire were quite similar with respect to the motivation question; however, the students did not prefer any one Proposal over the others in terms of wording.  Thus, it is clear that substantially shortening and simplifying Proposal C significantly improved students' understanding of the wording of the Proposal, while not reducing their understanding of the motivation.

\begin{figure}[htbp] 
\includegraphics[trim = 30mm 160mm 20mm 30mm, clip, scale = 0.7]{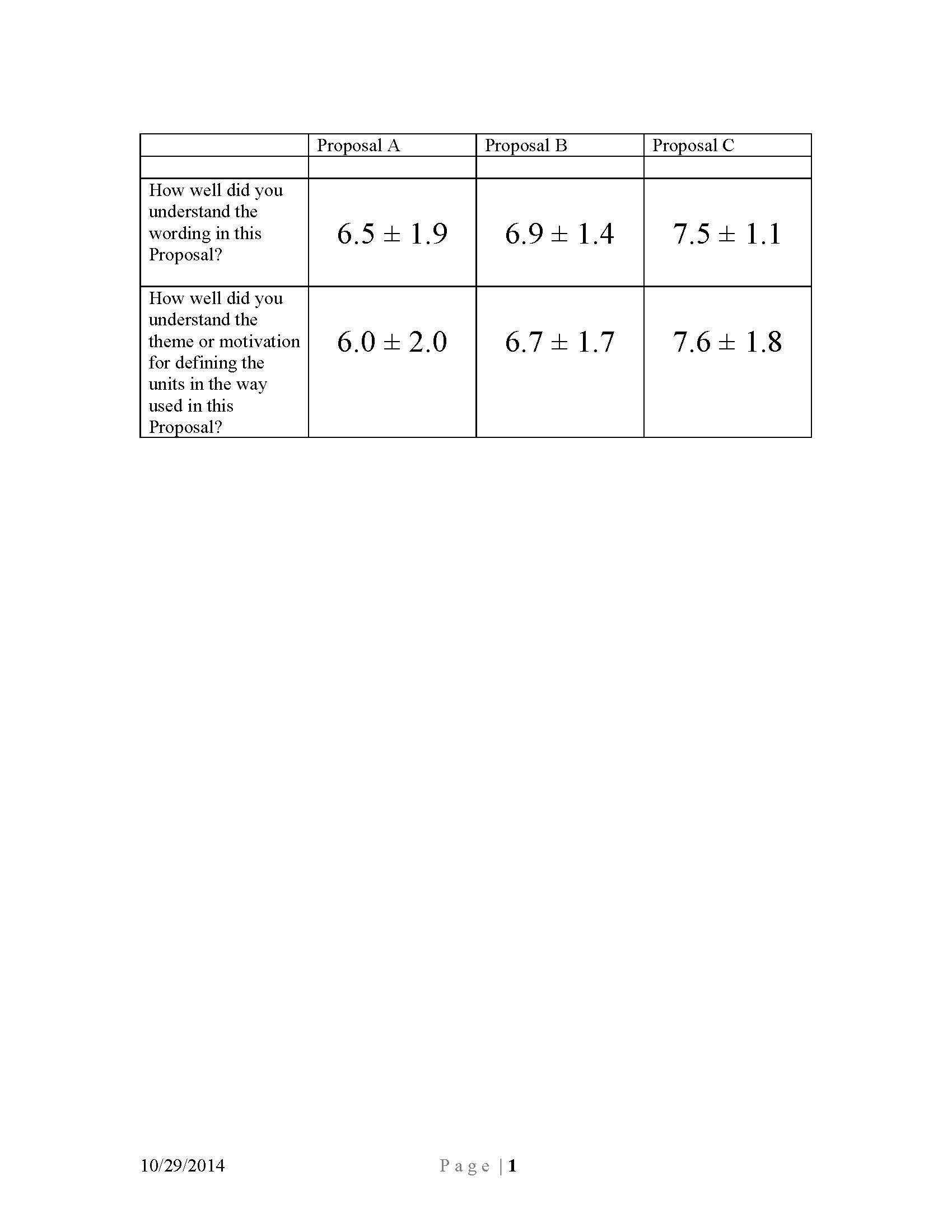}
\caption{\label{table} Questionnaire Numerical Results; the uncertainties are statistical (type A)
1$\sigma$ standard deviations.}
\end{figure}

\section{Conclusions and Next Steps}

As stated in the Introduction, our original motivation for pursuing this work was our
concern that, with the likely redefinition of the SI, students' initial introduction to
the subject might be more challenging and less straightforward.  Indeed, we believe that
the informal questionnaire that we offered (informal in the sense that we did not pursue a
detailed sampling algorithm, but simply let students self-select) bears out our concern:
students clearly indicated that the simple list of defined constants (Proposal A) is
significantly less clear in the motivation for defining a system of units in that way.

The questionnaire results also show that Proposal C in Figure \ref{fig:prop_C} satisfies
our two main goals: i) a clear conceptual underpinning for how to realize a system of
units, and ii) a simple exposition.

We do not insist that Proposal C is the only or the best possible way to introduce
students to the new SI.  However, we do believe that, for high school and undergraduate
teachers, it represents a useful, thought-provoking alternative to, for instance, the list
of defined constants embodied in Proposal A.  Our hope is that this paper, by pointing out
that the discussions in the new SI brochure are not the only possible ways to introduce
the new SI to students, helps to instigate a discussion amongst teachers, with the end
goal of providing the best introduction to students by both teachers and textbook authors.

We grateful acknowledge all of the students who spent their time responding to the questionnaires, Chris Lobb (University of Maryland) for both his encouragement and for facilitating the second questionnaire, and Bob Abel (Olympic College) for very helpful comments.

\section{Appendix: Questionnaire Cover Page}

\begin{figure}[htbp] 
\includegraphics[trim = 30mm 30mm 20mm 25mm, clip, scale = 0.7]{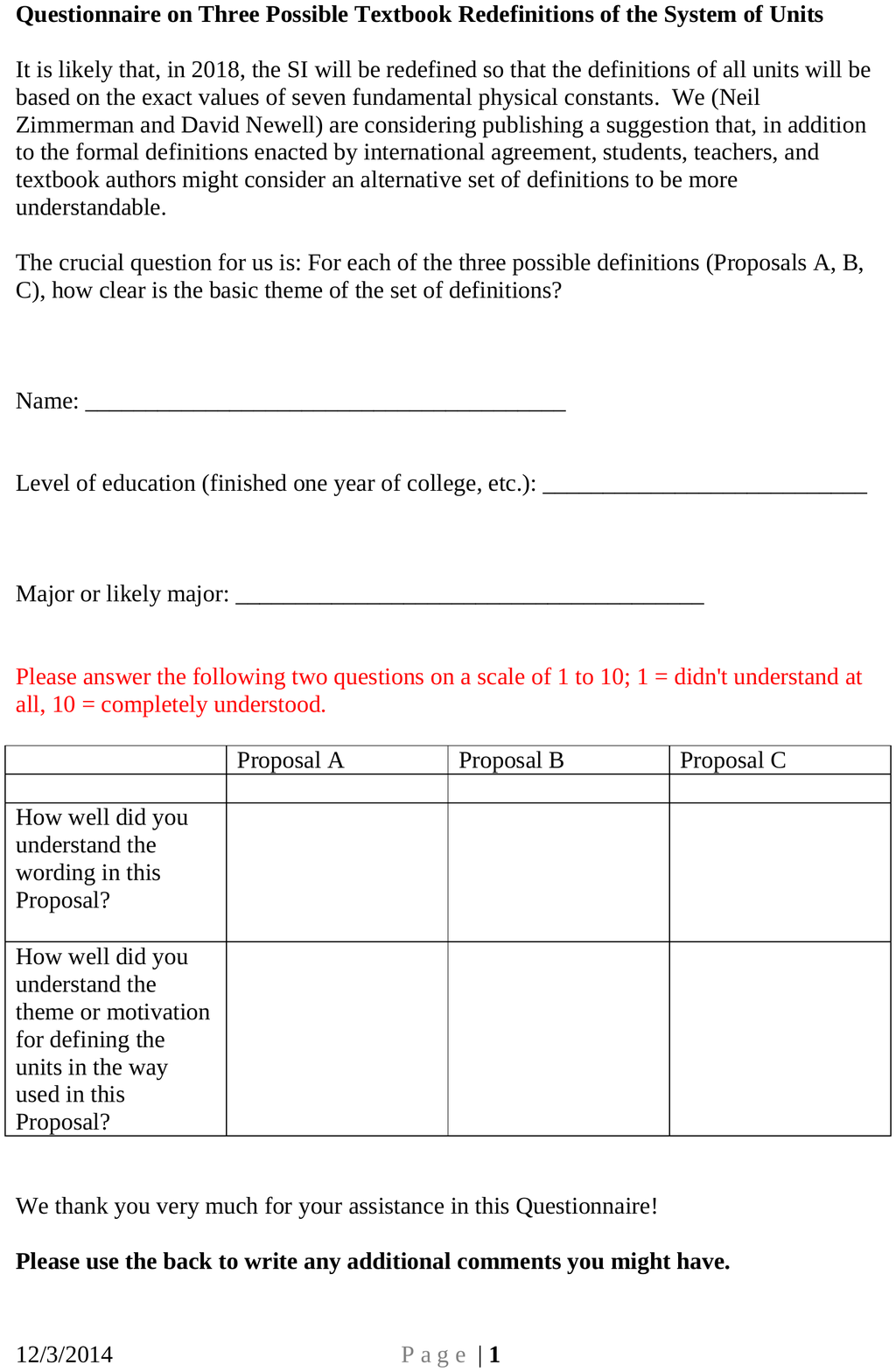}
\caption{\label{fig:cover_page} Cover Page of Questionnaire for Students}
\end{figure}

%
%
\bibliography{ref_units}

\begin{thebibliography}{9}
\expandafter\ifx\csname natexlab\endcsname\relax\def\natexlab#1{#1}\fi
\expandafter\ifx\csname bibnamefont\endcsname\relax
  \def\bibnamefont#1{#1}\fi
\expandafter\ifx\csname bibfnamefont\endcsname\relax
  \def\bibfnamefont#1{#1}\fi
\expandafter\ifx\csname citenamefont\endcsname\relax
  \def\citenamefont#1{#1}\fi
\expandafter\ifx\csname url\endcsname\relax
  \def\url#1{\texttt{#1}}\fi
\expandafter\ifx\csname urlprefix\endcsname\relax\def\urlprefix{URL }\fi
\providecommand{\bibinfo}[2]{#2}
\providecommand{\eprint}[2][]{\url{#2}}

\bibitem[{CIP(2013)}]{CIPM}
 (\bibinfo{year}{2013}), \bibinfo{note}{dec 2013 draft of the 9th SI Brochure,
  www.bipm.org/utils/common/pdf/si\_brochure\_draft\_ch123.pdf}.

\bibitem[{\citenamefont{Newell}(July, 2014)}]{Newell14a}
\bibinfo{author}{\bibfnamefont{D.~B.} \bibnamefont{Newell}},
  \bibinfo{journal}{Physics Today} pp. \bibinfo{pages}{35--40}
  (\bibinfo{year}{July, 2014}).

\bibitem[{\citenamefont{II}(2012)}]{Aubrecht12a}
\bibinfo{author}{\bibfnamefont{G.~J.~A.} \bibnamefont{II}},
  \bibinfo{journal}{The Physics Teacher} \textbf{\bibinfo{volume}{50}},
  \bibinfo{pages}{338 } (\bibinfo{year}{2012}).

\bibitem[{\citenamefont{Williams}(2014)}]{Williams14a}
\bibinfo{author}{\bibfnamefont{J.~H.} \bibnamefont{Williams}},
  \emph{\bibinfo{title}{{Defining and Measuring Nature: The Make of All
  Things}}} (\bibinfo{publisher}{Morgan and Claypool}, \bibinfo{address}{San
  Rafael, CA, USA}, \bibinfo{year}{2014}).

\bibitem[{\citenamefont{Adler}(2002)}]{Adler02a}
\bibinfo{author}{\bibfnamefont{K.}~\bibnamefont{Adler}},
  \emph{\bibinfo{title}{{The Measure of All Things}}}
  (\bibinfo{publisher}{Simon and Schuster}, \bibinfo{address}{New York, NY,
  USA}, \bibinfo{year}{2002}).

\bibitem[{\citenamefont{Pratt}(2014)}]{Pratt14a}
\bibinfo{author}{\bibfnamefont{J.~R.} \bibnamefont{Pratt}},
  \bibinfo{journal}{Measure} \textbf{\bibinfo{volume}{9(2)}},
  \bibinfo{pages}{26 } (\bibinfo{year}{2014}).

\bibitem[{\citenamefont{Fletcher et~al.}(2014)\citenamefont{Fletcher, Rietveld,
  Olthoff, Budovsky, and Milton}}]{Fletcher14a}
\bibinfo{author}{\bibfnamefont{N.}~\bibnamefont{Fletcher}},
  \bibinfo{author}{\bibfnamefont{G.}~\bibnamefont{Rietveld}},
  \bibinfo{author}{\bibfnamefont{J.}~\bibnamefont{Olthoff}},
  \bibinfo{author}{\bibfnamefont{I.}~\bibnamefont{Budovsky}}, \bibnamefont{and}
  \bibinfo{author}{\bibfnamefont{M.}~\bibnamefont{Milton}},
  \bibinfo{journal}{Measure} \textbf{\bibinfo{volume}{9(3)}},
  \bibinfo{pages}{30 } (\bibinfo{year}{2014}).

\bibitem[{\citenamefont{Mohr and Newell}(2010)}]{Mohr10a}
\bibinfo{author}{\bibfnamefont{P.~J.} \bibnamefont{Mohr}} \bibnamefont{and}
  \bibinfo{author}{\bibfnamefont{D.~B.} \bibnamefont{Newell}},
  \bibinfo{journal}{Am. J. Phys} \textbf{\bibinfo{volume}{78}},
  \bibinfo{pages}{338 } (\bibinfo{year}{2010}).

\bibitem[{\citenamefont{Zimmerman}(1998)}]{Zimmerman98a}
\bibinfo{author}{\bibfnamefont{N.~M.} \bibnamefont{Zimmerman}},
  \bibinfo{journal}{Am. J. Phys.} \textbf{\bibinfo{volume}{66}},
  \bibinfo{pages}{324} (\bibinfo{year}{1998}).

\end{thebibliography}

\clearpage

\end{document}